\documentclass[10pt,headsepline]{scrartcl}

\usepackage[latin1]{inputenc}   
\usepackage{latexsym}
\usepackage{amssymb}
\usepackage{amsfonts}
\usepackage{amsmath}
\usepackage[numbers,sort&compress]{natbib}
\usepackage{bm}
\usepackage{graphicx}
\usepackage{hyperref}
\usepackage{color}
\usepackage{multicol}

 \providecommand{\dprod}{\!\cdot\!}%
 \providecommand{\wprod}{\!\wedge \!}
 \providecommand{\pre}[1]{\,^#1}
 \providecommand{\sfrac}[2]{\frac{#1}{#2}\,}
 \providecommand{\email}[1]{email: \href{mailto:#1}{\texttt{#1}}}

\numberwithin{equation}{section}

\begin{document}

\newcommand{\mytitle}{The hidden geometric character of relativistic quantum mechanics}
\title{\mytitle}

\author{José B. Almeida\\ Universidade do Minho, Physics
Department\\
Braga, Portugal, \email{bda@fisica.uminho.pt}}

\date{}

\pagestyle{myheadings} \markright{José B. Almeida \hfill \mytitle}


\maketitle

\begin{abstract}
Geometry can be an unsuspected source of equations with physical
relevance, as everybody is aware since Einstein formulated the
general theory of relativity. However efforts to extend a similar
type of reasoning to other areas of physics, namely electrodynamics,
quantum mechanics and particle physics, usually had very limited
success; particularly in quantum mechanics the standard formalism is
such that any possible relation to geometry is impossible to detect;
other authors have previously trod the geometric path to quantum
mechanics, some of that work being referred to in the tex. In this
presentation we will follow an alternate route to show that quantum
mechanics has indeed a strong geometric character.

The paper makes use of geometric algebra, also known as Clifford
algebra, in 5-dimensional spacetime. The choice of this space is
given the character of first principle, justified solely by the
consequences that can be derived from such choice and their
consistency with experimental results. Given a metric space of any
dimension, one can define monogenic functions, the natural extension
of analytic functions to higher dimensions; such functions have null
vector derivative and have previously been shown by other authors to
play a decisive role in lower dimensional spaces.

All monogenic functions have null Laplacian by consequence; in an
hyperbolic space this fact leads inevitably to a wave equation with
plane-like solutions. This is also true for 5-dimensional spacetime
and we will explore those solutions, establishing a parallel with
the solutions of the free particle Dirac equation. For this purpose
we will invoke the isomorphism between the complex algebra of $4
\times 4$ matrices, also known as Dirac's matrices. There is one
problem with this isomorphism, because the solutions to Dirac's
equation are usually known as spinors (column matrices) that don't
belong to the $4 \times 4$ matrix algebra and as such are excluded
from the isomorphism. We will show that a solution in terms of Dirac
spinors is equivalent to a plane wave solution.

Just as one finds in the standard formulation, monogenic functions
can be naturally split into positive/negative energy together with
left/right ones. This split is provided by geometric projectors and
we will show that there is a second set of projectors providing an
alternate 4-fold split. The possible implications of this alternate
split are not yet fully understood and are presently the subject of
profound research.
\end{abstract}


\section{Introduction}
I have been advocating in recent papers that the majority of physics
equations can be derived from an appropriately chosen geometry by
exploration of the monogenic condition. Monogenic functions are not
familiar to everybody but they are really the natural extension of
analytic functions when one uses the formalism of geometric algebra
\cite{Doran03, Lasenby99, Hestenes84, Ryan04}; those functions zero
the vector derivative defined on the algebra of the particular
geometry under study.

In \cite{Almeida05:4} I showed how special relativity and the Dirac
equation could be derived from the monogenic condition applied in
the geometric algebra of 5-dimensional spacetime $G_{4,1}$. An
earlier paper \cite{Almeida05:1} proved that the same condition in
the same algebra was sufficient to produce a symmetry group
isomorphic to the standard model gauge group; unfortunately this
paper is incorrect in the formulation of particle dynamics but the
flaw was recently corrected;\cite{Almeida06:2} the same work
introduces electrodynamics and electromagnetism in the monogenic
formalism. Cosmological consequences were drawn from the addition of
an hyperspherical symmetry hypothesis with the consequent choice of
hyperspherical coordinates.\cite{Almeida05:3} Summing up all those
cited papers, I wrote a long book chapter;\cite{Almeida06:3} the
latest in the series is the derivation of energy states for the
hydrogen atom from the monogenic condition.\cite{Almeida06:1}

The present paper uses the 5D monogenic condition in 5D spacetime as
a postulate and explores its consequences for quantum mechanics,
establishing the conditions for equivalence to free particle Dirac's
equation but going beyond that and opening paths that may lead to
particle physics.

\section{Some geometric algebra \label{somealg}}
Geometric algebra is not usually taught in university courses and
its presence in the literature is scarce; good reference works are
\cite{Doran03, Hestenes84, Lasenby99}. We will concentrate on the
algebra of 5-dimensional spacetime because this will be our main
working space; this algebra incorporates as subalgebras those of the
usual 3-dimensional Euclidean space, Euclidean 4-space and Minkowski
spacetime. We begin with the simpler 5-D flat space and progress to
a 5-D spacetime of general curvature (see Appendix \ref{galgebra}
for more details.)

The geometric algebra ${G}_{4,1}$ of the hyperbolic 5-dimensional
space with signature $(-++++)$ is generated by the coordinate frame
of orthonormal basis vectors $\sigma_\alpha $ such that
\begin{eqnarray}
\label{eq:basis}
    && (\sigma_0)^2  = -1, \nonumber \\
    && (\sigma_i)^2 =1, \\
    && \sigma_\alpha \dprod \sigma_\beta  =0, \quad \alpha \neq \beta. \nonumber
    \nonumber
\end{eqnarray}
Note that the English characters $i, j, k$ range from 1 to 4 while
the Greek characters $\alpha, \beta, \gamma$ range from 0 to 4. See
Appendix \ref{indices} for the complete notation convention used.

Any two basis vectors can be multiplied, producing the new entity
called a bivector. This bivector is the geometric product or quite
simply the product, and it is distributive. Similarly to the product
of two basis vectors, the product of three different basis vectors
produces a trivector and so forth up to the fivevector, because five
is the spatial dimension.

We will simplify the notation for basis vector products using
multiple indices, i.e.\ $\sigma_\alpha \sigma_\beta \equiv
\sigma_{\alpha\beta}.$ The algebra is 32-dimensional and is spanned
by the basis
\begin{itemize}
\item 1 scalar, { $1$},
\item 5 vectors, { $\sigma_\alpha$},
\item 10 bivectors (area), { $\sigma_{\alpha\beta}$},
\item 10 trivectors (volume), { $\sigma_{\alpha\beta\gamma}$},
\item 5 tetravectors (4-volume), { $\mathrm{i} \sigma_\alpha $},
\item 1 pseudoscalar (5-volume), { $\mathrm{i} \equiv
\sigma_{01234}$}.
\end{itemize}
Several elements of this basis square to unity:
\begin{equation}
\label{eq:positive}
    (\sigma_i)^2 =  (\sigma_{0i})^2=
    (\sigma_{0i j})^2 =(\mathrm{i}\sigma_0)^2 =1.
\end{equation}
The remaining basis elements square to $-1$:
\begin{equation}
    \label{eq:negative}
    (\sigma_0)^2 = (\sigma_{ij})^2 = (\sigma_{ijk})^2 =
    (\mathrm{i}\sigma_i)^2 = \mathrm{i}^2=-1.
\end{equation}
Note that the pseudoscalar $\mathrm{i}$ commutes with all the other
basis elements while being a square root of $-1$; this makes it a
very special element which can play the role of the scalar imaginary
in complex algebra.

In 5-dimensional spacetime of general curvature, spanned by 5
coordinate frame vectors $g_\alpha$, the indices follow the
conventions set forth in Appendix \ref{indices}. We will also assume
this spacetime to be a metric space whose metric tensor is given by
\begin{equation}
\label{eq:metrictens}
    g_{\alpha \beta} = g_\alpha \dprod g_\beta;
\end{equation}
the double index is used with $g$ to denote the inner product of
frame vectors and not their geometric product. The space signature
is still $(-++++)$, which amounts to saying that $g_{00} < 0$ and
$g_{ii}
>0$. The coordinate frame vectors can be expressed as a linear
combination of the orthonormed ones, for which reason we have
\begin{equation}
    \label{eq:indexframemain}
    g_\alpha = {n^\beta}_\alpha \sigma_\beta,
\end{equation}
where ${n^\beta}_\alpha$ is called the \emph{refractive index
tensor} or simply the \emph{refractive index}; its 25 elements can
vary from point to point as a function of the
coordinates.\cite{Almeida04:4} In this work we will not consider
spaces of general curvature but only the connection-free ones, which
can be designated as bent spaces; in those spaces we define the
vector and covariant derivatives (see appendix \ref{derivatives}).

\section{Electromagnetism as gauge theory}
The simplest example of a gauge theory is electromagnetism, so we
will start by analysing this in the scope of monogenic functions.
The monogenic condition in flat space is given by\cite{Almeida06:3}
\begin{equation}
\label{eq:monogenic}
    \nabla \Psi = 0;
\end{equation}
which has plane wave like solutions of the type
\begin{equation}
\label{eq:planewave}
    \Psi = \psi_0 \mathrm{e}^{\mathrm{i}p_\alpha x^\alpha};
\end{equation}
where $p_0 = E$ is interpreted as energy, $p_4 = m$ as rest mass and
$\mathbf{p} = \sigma^m p_m$ as 3-dimensional momentum. Because the
second order equation $\nabla^2 \Psi = 0$ must also be verified, we
conclude easily that
\begin{equation}
    E^2 - \mathbf{p}^2 - m^2 = 0;
\end{equation}
that is $E \sigma^0 + p_m \sigma^m + m \sigma^4$ is a null vector.

A global symmetry of this equation is obtained with the
transformation
\begin{equation}
    \Psi' \mapsto \Psi \mathrm{e}^{\mathrm{i}\beta},
\end{equation}
where $\beta$ is a constant. It is easily verified that if $\Psi$ is
monogenic so is $\Psi'$ and the symmetry is global because $\beta$
is the same everywhere; the quantity $\exp(\mathrm{i}\beta)$ is a
phase factor. If we allow $\beta$ to change with spacetime
coordinates, $\beta = \beta(x^\mu)$, the monogenic condition will no
longer be verified by $\Psi'$
\begin{equation}
\label{eq:nonmonogenic}
    \nabla \Psi' = \nabla \Psi \mathrm{e}^{\mathrm{i}\beta} +
    \mathrm{i}\pre{\mu}\nabla \beta \Psi \mathrm{e}^{\mathrm{i}\beta}.
\end{equation}
We can define a local symmetry by changing the monogenic condition
through replacement of the vector derivative by a covariant
derivative which cancels out the extra term in Eq.\
(\ref{eq:nonmonogenic}); we do this as usual, by defining a vector
field $A = \sigma^\mu A_\mu$ and writing the covariant derivative as
\begin{equation}
\label{eq:emderiv}
    \mathrm{D} = \pre{\mu}\nabla + \left(\sigma^4 + \sfrac{q}{m} A \right) \partial_4;
\end{equation}
where $A_\mu$ transforms as
\begin{equation}
    A' \mapsto A - \sfrac{1}{q} \pre{\mu}\nabla \beta.
\end{equation}
In the two equations above $q$ and $m$ are charge and mass
densities, respectively; in the non-dimensional units system $q=-1$
for an electron. Applying the extended monogenic condition to
$\Psi'$ we get
\begin{equation}
    \mathrm{D}' \Psi' = \nabla \Psi \mathrm{e}^{\mathrm{i}\beta} +
    \mathrm{i} q A \Psi \mathrm{e}^{\mathrm{i}\beta} = \mathrm{D} \Psi
    \mathrm{e}^{\mathrm{i}\beta}.
\end{equation}
So, obviously, if $\mathrm{D}\Psi$ is null, that is if $\Psi$ is
monogenic in an extended sense, so is $\Psi'$. Now, the covariant
derivative in Eq.\ (\ref{eq:emderiv}) can be associated with the
reciprocal frame
\begin{equation}
\label{eq:em_frame}
    g^\mu = \sigma^\mu,~~~~ g^4 = \sigma^4 + \sfrac{q}{m}A_\mu \sigma^\mu,
\end{equation}
which was introduced in Refs.\ \cite{Almeida06:2, Almeida06:3} to
derive electrodynamics and electromagnetism from the monogenic
condition. We see then that electromagnetic gauge invariance is a
consequence of a non-orthonormed $g^4$ frame vector and can be
accommodated in the refractive index formalism introduced above.

\section{The Dirac equation}
In this section we examine the monogenic condition
(\ref{eq:monogenic}) and the solution (\ref{eq:planewave}) to
establish the conditions under which they become equivalent to free
particle Dirac's equation; the section ends with an indication of
the procedure for the inclusion of an EM field. We will accept
without explanation that the solution $\Psi$ has harmonic dependence
on $x^4$ with a frequency equal to a particle's rest mass and write
\begin{equation}
\label{eq:restmass}
    \Psi = \psi \mathrm{e}^{\mathrm{i}m x^4}.
\end{equation}
At the present moment we cannot offer a plausible explanation for
the existence of eigenvalues for the rest mass, corresponding to the
various elementary particles; this is still a mystery with which we
must live if we want to proceed. We note, however, that no existing
theory offers a satisfactory explanation for the observed elementary
particles' masses and so we are no worse than all other theories in
that respect.

When the monogenic condition (\ref{eq:monogenic}) is applied to
$\Psi$ with the assumption above we get
\begin{equation}
\label{eq:masseq}
    (\sigma^0 \partial_t + \sigma^m \partial_m + \mathrm{i}
    \sigma^4 m) \psi = 0;
\end{equation}
multiplying by $\mathrm{i}\sigma^0$ on the left
\begin{equation}
\label{eq:monogenic_2}
    (\mathrm{i}\partial_t + \mathrm{i}\sigma^{m0} \partial_m -
    \sigma^{40} m ) \psi = 0.
\end{equation}
Geometric algebra $G_{4,1}$ is isomorphic to the complex algebra of
$4 \ast 4$ matrices \cite{Lounesto01}, which means we can represent
by matrix equations all the geometric equations in this paper. By
resorting to matrix equations we will loose the geometric content
inherent to geometric algebra; nevertheless such step is crucial if
we want to understand the relationship with Dirac's equation because
the latter uses matrix formalism. In order to make the transition be
tween the two equations we need a matrix representation for
geometric algebra elements; although there are many possible choices
we will adopt the Dirac-Pauli representation, represented in Table
\ref{t:Dirac}, because it is commonly found in the literature.
\begin{table*}
\caption{\label{t:Dirac} Dirac-Pauli matrices.}
\begin{alignat*}{2}
    \alpha^1 &=   \begin{pmatrix} 0 & 0 & 0 & 1  \\
    0 & 0 & 1 & 0 \\ 0 & 1 & 0 & 0 \\ 1 & 0 & 0 &
    0
    \end{pmatrix},&~
    \alpha^2 &=  \begin{pmatrix} 0 & 0 & 0 & -\mathrm{i} \\
    0 & 0 & \mathrm{i} & 0 \\ 0 & -\mathrm{i} & 0 & 0 \\ \mathrm{i} & 0 & 0 & 0
    \end{pmatrix},\\
    \alpha^3 &=   \begin{pmatrix} 0 & 0 & 1 & 0 \\
    0 & 0 & 0 & -1 \\ 1 & 0 & 0 & 0 \\ 0 & -1 & 0 & 0
    \end{pmatrix},&~
    \beta &= \begin{pmatrix} 1 & 0 & 0 & 0 \\
    0 & 1 & 0 & 0 \\ 0 & 0 & -1 & 0 \\ 0 & 0 & 0 & -1
    \end{pmatrix} .
\end{alignat*}
\end{table*}
We can now associate geometric algebra elements to their matrix
counterparts
\begin{equation}
    \sigma^{m0} \equiv \alpha^m,~~~~ \sigma^{40} \equiv \beta.
\end{equation}
The different basis vectors of geometric algebra can then be given
matrix equivalents by
\begin{equation}
\label{eq:matrixequiv}
\begin{split}
    {\sigma^0} &\equiv {\mathrm{i} \alpha^1 \alpha^2
    \alpha^3 \beta },\\
    {\sigma^1} &\equiv {\mathrm{i} \alpha^2
    \alpha^3 \beta},\\
    {\sigma^2} &\equiv {\mathrm{i}
    \alpha^1\alpha^3 \beta},\\
    {\sigma^3} &\equiv {-\mathrm{i} \alpha^1
    \alpha^2 \beta},\\
    {\sigma^4} &\equiv {\mathrm{i}\alpha^1 \alpha^2
    \alpha^3};
\end{split}
\end{equation}
resulting in the matrices of Table \ref{t:sigma}
\begin{table*}
\caption{\label{t:sigma} Matrix representation of basis vectors}
\begin{alignat*}{3}
    {\sigma^0} &\equiv   {\begin{pmatrix} 0 & 0 & 1 & 0  \\
    0 & 0 & 0 & 1 \\ -1 & 0 & 0 & 0 \\ 0 & -1 & 0 &    0
    \end{pmatrix}}, &~~
    {\sigma^1} &\equiv  {\begin{pmatrix} 0 & 1 & 0 & 0 \\
    1 & 0 & 0 & 0 \\ 0 & 0 & 0 & -1 \\ 0 & 0 & -1 & 0
    \end{pmatrix}}, &~~
    {\sigma^2} &\equiv   {\begin{pmatrix} 0 & -\mathrm{i} & 0 & 0 \\
    \mathrm{i} & 0 & 0 & 0 \\ 0 & 0 & 0 & \mathrm{i} \\ 0 & 0 & -\mathrm{i} & 0
    \end{pmatrix}}, \\
    {\sigma^3} &\equiv {\begin{pmatrix} 1 & 0 & 0 & 0 \\
    0 & -1 & 0 & 0 \\ 0 & 0 & -1 & 0 \\ 0 & 0 & 0 & 1
    \end{pmatrix}},&~~
    {\sigma^4} &\equiv   {\begin{pmatrix} 0 & 0 & -1 & 0 \\
    0 & 0 & 0 & -1 \\ -1 & 0 & 0 & 0 \\ 0 & -1 & 0 & 0
    \end{pmatrix}} .
\end{alignat*}
\end{table*}

The monogenic condition (\ref{eq:monogenic}) can then be written in
the alternative form
\begin{equation}
\label{eq:Dirac_2}
    (\mathrm{i} \partial_t + \mathrm{i} \alpha^m
    \partial_m
    - \beta m) \psi = 0;
\end{equation}
this can be read either as a geometric or matrix equation. In the
latter case one recognizes that it has become Dirac's own equation
in the absence of fields. There are problems however in associating
the plane wave solutions given by Eq.\ (\ref{eq:planewave}) to those
of Dirac's equation, mainly because in the geometric case the
solutions will have $4 \ast 4$ matrix representations while the
spinors associated with Dirac's solutions are column matrices which
lay outside the matrix algebra. In order to bring Dirac's spinors
into the matrix algebra we construct a matrix by placing the four
solutions to Dirac's equation side by side
\begin{equation}
    \bar{\psi} = \begin{pmatrix}
    \bar{\psi}_1 &
    \bar{\psi}_2 &
    \bar{\psi}_3 &
    \bar{\psi}_4
    \end{pmatrix};
\end{equation}
where $\bar{\psi}_i$ are the four eigenvectors verifying the Dirac
equation. The overbar is used, where necessary, to indicate Dirac
equation elements that must be represented in their matrix form.
Dirac's equation can now be written in its eigenvalue form
\begin{equation}
    \bar{A}\bar{\psi} = \bar{\psi}\bar{\Lambda},
\end{equation}
where $\bar{A}$ is the matrix representation of $\alpha^m p_m  +
\beta m$ and $\bar{\Lambda}$ is a diagonal matrix containing the
eigenvalues of $\bar{A}$. This form of Dirac's equation is
equivalent to four instances of the usual free particle equation,
each with its own spinor solution; the non-zero elements in
$\Lambda$ are obviously $\pm E$. From the equation above one
concludes easily that matrix $\bar{A}$ can be decomposed as
\begin{equation}
\label{eq:decompose}
    \bar{A} = \bar{\psi}\bar{\Lambda}\bar{\psi}^{-1}.
\end{equation}

We return now to Eq.\ (\ref{eq:masseq}) to verify what it implies
for $\psi$. The first member contains the product of a null vector
by some function, which must ultimately be null; this implies that
$\psi$ must contain a null vector factor $E \sigma^0 + p_m \sigma^m
+ m \sigma^4$, although other factors are not excluded. We then make
the new assumption that $\psi$ is of the form
\begin{equation}
\begin{split}
\label{eq:psi0}
    \psi &= (-E + p_m \sigma^{m0} + m \sigma^{40})\mathrm{e}^{\mathrm{i} (-E t
    + p_m x^m)} \\
    &\equiv (- E \bar{I} + \bar{A})\mathrm{e}^{\mathrm{i} (-E t
    + p_m x^m)}.
\end{split}
\end{equation}
Inserting into Eq.\ (\ref{eq:Dirac_2}) we see that the following
must be verified
\begin{equation}
    \bar{A} (E\bar{I} + \bar{A}) = E (E\bar{I} + \bar{A});
\end{equation}
and so it must be
\begin{equation}
    \bar{A}^2 = E^2 \bar{I}.
\end{equation}
Replacing one of the $\bar{A}$ on the left hand side with Eq.\
(\ref{eq:decompose})
\begin{equation}
    \bar{A} \bar{\psi}\bar{\Lambda} \bar{\psi}^{-1} = E^2 \bar{I} .
\end{equation}
Multiplying both sides on the right by $\bar{\psi}
\bar{\Lambda}^{-1}$
\begin{equation}
    \bar{A}\bar{\psi} = E^2 \bar{\psi} \bar{\Lambda}^{-1}.
\end{equation}
But $E^2\bar{\Lambda}^{-1} = \bar{\Lambda}$ and so the monogenic
equation is equivalent to the eigenvalue equation
\begin{equation}
\label{eq:Dirac_3}
    \bar{A} \bar{\psi} = \bar{\psi} \bar{\Lambda}.
\end{equation}
Remember that $\bar{\psi}$ is a matrix containing the four
eigenvectors of $\bar{A}$ and $\bar{\Lambda}$ is diagonal with the
eigenvalues of $\bar{A}$, so we have just written four instances of
Dirac's equation in matrix form.

A geometric equivalent to Dirac´s equation for a particle in an
electromagnetic field can be immediately obtained by extension of
the monogenic condition (\ref{eq:monogenic}). If the derivative
operator $\nabla$ is replaced by $\mathrm{D}$ from Eq.\
(\ref{eq:emderiv}), the new geometric condition represents particles
in fields, in the same way as Dirac's equation; this was proven for
the case of the hydrogen atom in \cite{Almeida06:1}. We say that we
are applying an extended monogenic condition because the original
monogenic condition is applied in a curved space with frame vectors
given by Eq.\ (\ref{eq:em_frame}).

\section{Energy and helicity splits}
When we imposed that $\bar{\psi}$ should have the four eigenvectors
of $\bar{A}$ side by side we said nothing about ordering them
because different ordering options only imply rearrangements of
positive and negative elements in the diagonal matrix
$\bar{\Lambda}$. We are then free to choose one of the possible
orders that lead to
\begin{equation}
    \bar{\Lambda} = E \beta = E \sigma^{40}.
\end{equation}
It turns out that $(1 + \sigma^{40})/2$ and $(1 - \sigma^{40})/2$
are two orthogonal idempotents of $G_{4,1}$. This means that they
are both elements that square to themselves, thus the idempotent
designation, and their product is zero. In matrix representation,
they are both diagonal matrices, the former with the top two
elements equal to identity and the bottom pair equal to zero and the
latter with the positions reversed. It is then clear that by
multiplying Eq.\ (\ref{eq:Dirac_3}) on the right by one of the
idempotents we can select either the positive or negative energy
equations from the set of four equations represented in the original
form.

It is actually possible to define sets of four orthogonal
idempotents in $G_{4,1}$ in different ways; one such possibility
results from multiplying each of the energy split idempotents just
defined by one of the pair  $(1 + \sigma^3)/2$ and $(1 -
\sigma^3)/2$. These idempotents select left or right spin solutions
and we call this the helicity split. The set of four idempotents is
then
\begin{equation}
\begin{split}
    f_1 &= \frac{1}{4}\, (1- \sigma^3)(1 - \sigma^{04});\\
    f_2 &= \frac{1}{4}\, (1- \sigma^3)(1 + \sigma^{04});\\
    f_3 &= \frac{1}{4}\, (1+ \sigma^3)(1 + \sigma^{04});\\
    f_4 &= \frac{1}{4}\, (1+ \sigma^3)(1 - \sigma^{04}).
\end{split}
\end{equation}
Another set of four orthogonal idempotents is provided by the
definitions
\begin{equation}
\begin{split}
    e_1 &= \frac{1}{4}\, (1+ \sigma^{012})(1 + \sigma^{034});\\
    e_2 &= \frac{1}{4}\, (1+ \sigma^{012})(1 - \sigma^{034});\\
    e_3 &= \frac{1}{4}\, (1- \sigma^{012})(1 - \sigma^{034});\\
    e_4 &= \frac{1}{4}\, (1- \sigma^{012})(1 + \sigma^{034}).
\end{split}
\end{equation}
This set is not totally independent from the previous one, because
the $\sigma^{034}$ element is present in both sets, but the two sets
don't obviously overlap. In order to get some insight into the
physical meaning of this second set we will have to look at some of
the symmetries in $G_{4,1}$.

\begin{table*}[htb]
\caption{\label{t:lambda} Generators for $SU(4)$ symmetry group}
\begin{alignat*}{3}
\label{eq:gelmann}
    \lambda_1 &\equiv  \begin{pmatrix} 0 & 1 & 0 & 0 \\
    1 & 0 & 0 & 0 \\ 0 & 0 & 0 & 0 \\ 0 & 0 & 0 & 0
    \end{pmatrix}, &~~
    \lambda_2 &\equiv  \begin{pmatrix} 0 & -\mathrm{i} & 0 & 0 \\
    \mathrm{i} & 0 & 0 & 0 \\ 0 & 0 & 0 & 0 \\ 0 & 0 & 0 & 0
    \end{pmatrix}, &~~
    \lambda_3 &\equiv  \begin{pmatrix} 1 & 0 & 0 & 0 \\
    0 & -1 & 0 & 0 \\ 0 & 0 & 0 & 0 \\ 0 & 0 & 0 & 0
    \end{pmatrix},\\
    \lambda_4 &\equiv  \begin{pmatrix} 0 & 0 & 0 & 0 \\
    0 & 0 & 1 & 0 \\ 0 & 1 & 0 & 0 \\ 0 & 0 & 0 & 0
    \end{pmatrix}  ,  &~~
    \lambda_5 &\equiv  \begin{pmatrix} 0 & 0 & 0 & 0 \\
    0 & 0 & -\mathrm{i} & 0 \\ 0 & \mathrm{i} & 0 & 0 \\ 0 & 0 & 0 & 0
    \end{pmatrix}, &~~
    \lambda_6 &\equiv  \begin{pmatrix} 0 & 0 & 1 & 0 \\
    0 & 0 & 0 & 0 \\ 1 & 0 & 0 & 0 \\ 0 & 0 & 0 & 0
    \end{pmatrix}  \\
    \lambda_7 &\equiv  \begin{pmatrix} 0 & 0 & -\mathrm{i} & 0 \\
    0 & 0 & 0 & 0 \\ 0 & \mathrm{i} & 0 & 0 \\ 0 & 0 & 0 & 0
    \end{pmatrix},   &~~
    \lambda_8 &\equiv \frac{1}{\sqrt{3}} \begin{pmatrix} 1 & 0 & 0 & 0 \\
    0 & 1 & 0 & 0 \\ 0 & 0 & -2 & 0 \\ 0 & 0 & 0 & 0
    \end{pmatrix}  , &~~
    \lambda_9 &\equiv  \begin{pmatrix} 0 & 0 & 0 & 1 \\
    0 & 0 & 0 & 0 \\ 0 & 0 & 0 & 0 \\ 1 & 0 & 0 & 0
    \end{pmatrix}, \\
    \lambda_{10} &\equiv  \begin{pmatrix} 0 & 0 & 0 & -\mathrm{i} \\
    0 & 0 & 0 & 0 \\ 0 & 0 & 0 & 0 \\ \mathrm{i} & 0 & 0 & 0
    \end{pmatrix}, &~~
    \lambda_{11} &\equiv  \begin{pmatrix} 0 & 0 & 0 & 0 \\
    0 & 0 & 0 & 1 \\ 0 & 0 & 0 & 0 \\ 0 & 1 & 0 & 0
    \end{pmatrix}, &~~
    \lambda_{12} &\equiv  \begin{pmatrix} 0 & 0 & 0 & 0 \\
    0 & 0 & 0 & -\mathrm{i} \\ 0 & 0 & 0 & 0 \\ 0 & \mathrm{i} & 0 & 0
    \end{pmatrix}  ,  \\
    \lambda_{13} &\equiv  \begin{pmatrix} 0 & 0 & 0 & 0 \\
    0 & 0 & 0 & 0 \\ 0 & 0 & 0 & 1 \\ 0 & 0 & 1 & 0
    \end{pmatrix}, &~~
    \lambda_{14} &\equiv  \begin{pmatrix} 0 & 0 & 0 & 0 \\
    0 & 0 & 0 & 0 \\ 0 & 0 & 0 & -\mathrm{i} \\ 0 & 0 & \mathrm{i} & 0
    \end{pmatrix},  &~~
    \lambda_{15} &\equiv  \frac{1}{\sqrt{6}} \begin{pmatrix} 1 & 0 & 0 & 0 \\
    0 & 1 & 0 & 0 \\ 0 & 0 & 1 & 0 \\ 0 & 0 & 0 & -3
    \end{pmatrix} .
\end{alignat*}
\end{table*}

The isomorphism between geometric algebra $G_{4,1}$ and matrix
algebra $M(4,C)$ allows us to state that unitary elements of the
algebra are elements of symmetry group $U(4)$ and those with unit
determinant are elements of symmetry group $SU(4)$. We don't need
any further demonstration because we can always find a geometric
algebra equivalent to any matrix element of $SU(4)$, however the
result of this conversion is dependent on the particular assignment
made between basis vectors and matrices and there are infinitely
many ways of making such assignment. All elements of an $SU(4)$
group can be generated by a set of 15 generators; in matrix form,
one frequent choice is shown in Table \ref{t:lambda}.\cite{Greiner01} 
One verifies easily that there are 3 diagonal elements:
$\lambda_3$, $\lambda_8$ and $\lambda_{15}$; since the four $f_i$
idempotents have diagonal matrix representations, they can be
written as linear combinations of the 3 diagonal generators and
unit, as follows
\begin{equation}
\begin{split}
    f_1 &= \frac{1}{4} { + \frac{\lambda_3}{2} + \frac{\lambda_8}{2
    \sqrt{3}}} + \frac{\lambda_{15}}{2 \sqrt{6}}; \\
    f_2 &= \frac{1}{4} { - \frac{\lambda_3}{2} + \frac{\lambda_8}{2
    \sqrt{3}}} + \frac{\lambda_{15}}{2 \sqrt{6}}; \\
    f_3 &= \frac{1}{4} { - \frac{\lambda_8}{\sqrt{3}}} + \frac{\lambda_{15}}{2 \sqrt{6}}; \\
    f_4 &= \frac{1}{4} - \sqrt{\frac{3}{2}}\,\lambda_{15}.
\end{split}
\end{equation}
The relations above would be equally true if we had chosen a
different matrix assignment for basis vectors such that the
idempotents were no longer represented by diagonal matrices. If we
analyse the set of $e_i$ idempotents, it is clear that they are not
represented by diagonal matrices but this is only circunstancial for
we could have chosen a matrix assignment that would render $e_i$
diagonal instead of $f_i$. This is to conclude that any orthogonal
set of 4 idempotents can be represented as linear combinations of
privileged $SU(4)$ generators; reversing the relations above one
gets
\begin{equation}
\begin{split}
    \lambda_3 &= f_1 - f_2,\\
    \lambda_8 &= \sfrac{f_1 + f_2 - 2 f_3}{\sqrt{3}},\\
    \lambda_{15} &= \sfrac{f_1+f_2+f_3-3f_4}{\sqrt{6}}.
\end{split}
\end{equation}
To end this subject we will just say that $e_i$ idempotents may
encode isospin, strangeness and charm in a still unclear way.
\citet{Greiner01} encode these quantum numbers into the diagonal
generators of $SU(4)$, which gives some substance to our claim.

\section{Stable wavepackets}
The fact that particles appear with stable spatial distributions
seems incompatible with a wavepacket description. It is widely known
that even a Gaussian beam spreads out, in spite of remaining
Gaussian; other distributions will not even conserve shape.
\citet{Bohm05} suggested that a Gaussian distribution could be
stabilized by influence of quantum vacuum but we will show next that
the monogenic condition can produce naturally stable wavepackets
without appealing to foreign mechanisms.

Following the reasoning in previous paragraphs, we expect a
stationary particle to be described by a wavefunction of the type
\begin{equation}
    \Psi = \psi_r(x^m) \psi_t(x^0,x^4);
\end{equation}
where $\psi_r$ describes the spatial distribution and $\psi_t$
describes the propagation along $x^4$. Applying the monogenic
condition we get
\begin{equation}
    \nabla \psi_r \psi_t + \overrightarrow{\nabla}\psi_r  \overleftarrow{\psi}_t =0.
\end{equation}
This equation is verified if each term on the left hand side is
independently equal to zero. In the first term we note that $\nabla
\psi_r = 0$ defines 3-dimensional monogenic functions or spherical
harmonics, fully described in \citet{Doran03}. These functions have
only scalar and bivector $\sigma^{12}$ terms and so we can argue
that $\nabla$ in the second term commutes with $\psi_r$. This is
true because the only derivatives to be considered are with respect
to $x^0$ and $x^4$ and both $\sigma^0$ and $\sigma^4$ will commute
with $\psi_r$. We are then led to solve $\nabla \psi_t = 0$, which
produces wave solutions propagating along $x^4$. We have then
established several stable spatial distributions, propagating along
$x^4$, compatible with stationary particles. For non-stationary
particles we can rotate the solutions in 4D, maintaining stable
spatial distributions. Although we don't have yet a clear
interpretation of this possibility, we believe it is likely that it
will help understand why particles exhibit stable spatial
distributions.

\section{Conclusion and future work}
The use of 5-dimensional spacetime and monogenic functions as first
principles allows the derivation of numerous equations of high
physical significance; in this paper we explored some of those with
relevance for quantum mechanics. The widely accepted principle of
gauge invariance was shown to be fully equivalent to 5D space
curvature, allowing the formulation of Maxwell's equations and
electrodynamics in the framework of monogenic functions.

In respect to Dirac's equation our point of departure was the
monogenic condition in 5D spacetime expressed by Eq.\
(\ref{eq:monogenic}), which has plane wave type solutions. We then
added the constraint that $x^4$ dependence should be harmonic,
involving the particle's rest mass and we expressed that constraint
by Eq.\ (\ref{eq:restmass}). The constant factor in plane wave
solutions of the monogenic condition must include a null vector but
it can include other factors; we added the constraint that there was
also a $\sigma^0$ factor, resulting in the form given by Eq.\
(\ref{eq:psi0}). Those two constraints are enough to render the
monogenic condition fully equivalent to free particle Dirac's
equation in its conventional form; we can then say that the
monogenic condition fully includes Dirac's equation but it allows
more general solutions which may or may not have physical
significance.

It was shown that any set of four orthogonal idempotents is closely
related to the diagonal generators of $SU(4)$ in a particular matrix
representation. A second set of four orthogonal idempotents was
identified, independent of the energy/helicity set; it can be argued
that this second set is related to the standard model gauge group of
quarks bosons and hadrons. This is little more than a speculation at
present and more work is needed in order to fully understand the
meaning of the two superimposed symmetries. The paper ends with a
demonstration that the monogenic condition produces non-expanding 4D
wavepackets that may become very useful for modelling stable
particles.
\begin{figure*}[htb]
\vspace{11pt} \centerline{\includegraphics[scale=1]{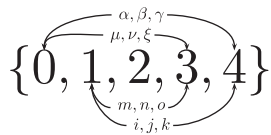}}

\caption{\label{f:indices} Indices in the range $\{0,4\}$ will be
denoted with Greek letters $\alpha, \beta, \gamma.$ Indices in the
range $\{0,3\}$ will also receive Greek letters but chosen from
$\mu, \nu, \xi.$ For indices in the range $\{1,4\}$ we will use
Latin letters $i, j, k$ and finally for indices in the range
$\{1,3\}$ we will use also Latin letters chosen from $m, n, o.$ }
\end{figure*}

\begin{appendix}
\section{Indexing conventions \label{indices}}
In this section we establish the indexing conventions used in the
paper. We deal with 5-dimensional space but we are also interested
in two of its 4-dimensional subspaces and one 3-dimensional
subspace; ideally our choice of indices should clearly identify
their ranges in order to avoid the need to specify the latter in
every equation. The diagram in Fig.\ \ref{f:indices} shows the index
naming convention used in this paper; Einstein's summation
convention will be adopted as well as the compact notation for
partial derivatives $\partial_\alpha =
\partial/\partial x^\alpha.$

\section{Non-dimensional units \label{units}}
\begin{table*}[htb]
\caption{\label{t:standards}Standards for non-dimensional units'
system}
\begin{center}
\begin{tabular}{c|c|c|c}
Length & Time & Mass & Charge \\
\hline & & & \\

$\displaystyle \sqrt{\frac{G \hbar}{c^3}} $ & $\displaystyle
\sqrt{\frac{G \hbar}{c^5}} $  & $\displaystyle \sqrt{\frac{ \hbar c
}{G}} $  & $e$
\end{tabular}
\end{center}
\end{table*}
The interpretation of $t$ and $\tau$ as time coordinates implies the
use of a scale parameter which is naturally chosen as the vacuum
speed of light $c$. We don't need to include this constant in our
equations because we can always recover time intervals, if needed,
introducing the speed of light at a later stage. We can even go a
step further and eliminate all units from our equations so that they
become pure number equations; in this way we will avoid cumbersome
constants whenever coordinates have to appear as arguments of
exponentials or trigonometric functions. We note that, at least for
the macroscopic world, physical units can all be reduced to four
fundamental ones; we can, for instance, choose length, time, mass
and electric charge as fundamental, as we could just as well have
chosen others. Measurements are then made by comparison with
standards; of course we need four standards, one for each
fundamental unit. But now note that there are four fundamental
constants: Planck constant $(\hbar)$, gravitational constant $(G)$,
speed of light in vacuum $(c)$ and proton electric charge $(e)$,
with which we can build four standards for the fundamental units.

Table \ref{t:standards} lists the standards of this units' system,
frequently called Planck units, which the authors prefer to
designate by non-dimensional units. In this system all the
fundamental constants, $\hbar$, $G$, $c$, $e$, become unity, a
particle's Compton frequency, defined by $\nu = mc^2/\hbar$, becomes
equal to the particle's mass and the frequent term ${GM}/({c^2 r})$
is simplified to ${M}/{r}$. We can, in fact, take all measures to be
non-dimensional, since the standards are defined with recourse to
universal constants; this will be our posture. Geometry and physics
become relations between pure numbers, vectors, bivectors, etc. and
the geometric concept of distance is needed only for graphical
representation.

\section{Some complements of geometric algebra \label{galgebra}}
In this section we expand the concepts given in Sec.\ \ref{somealg},
introducing some useful relations and definitions. The geometric
product of any two vectors $a = a^\alpha \sigma_\alpha$ and $b =
b^\beta \sigma_\beta$ can be found making use of the distributive
property and the already defined products of basis vectors
\begin{equation}
    ab = \left(-a^0 b^0 + \sum_i a^i b^i \right) + \sum_{\alpha \neq \beta}
    a^\alpha b^\beta \sigma_{\alpha \beta}.
\end{equation}
We notice it can be decomposed into a symmetric part, a scalar
called the inner or interior product, and an anti-symmetric part, a
bivector called the outer or exterior product.
\begin{equation}
    ab = a \dprod b + a \wprod b,~~~~ ba = a \dprod b - a \wprod b.
\end{equation}
Reversing the definition one can write inner and outer products as
\begin{equation}
    a \dprod b = \frac{1}{2}\, (ab + ba),~~~~ a \wprod b = \frac{1}{2}\, (ab -
    ba).
\end{equation}
The inner product is the same as the usual ''dot product,'' the only
difference being in the negative sign of the $a_0 b_0$ term; this is
to be expected and is similar to what one finds in special
relativity. The outer product represents an oriented area; in
Euclidean 3-space it can be linked to the "cross product" by the
relation $\mathrm{cross}(\mathbf{a},\mathbf{b}) = - \sigma_{123}
\mathbf{a} \wprod \mathbf{b}$; here we introduced bold characters
for 3-dimensional vectors and avoided defining a symbol for the
cross product because we will not use it again. We also used the
convention that interior and exterior products take precedence over
geometric product in an expression.

When a vector is operated with a multivector the inner product
reduces the grade of each element by one unit and the outer product
increases the grade by one. We will generalize the definition of
inner and outer products below; under this generalized definition
the inner product between a vector and a scalar produces a vector.
Given a multivector $a$ we refer to its grade-$r$ part by writing
$<\!a\!>_r$; the scalar or grade zero part is simply designated as
$<\!a\!>$. By operating a vector with itself we obtain a scalar
equal to the square of the vector's length
\begin{equation}
    a^2 = aa = a \dprod a + a \wprod a = a \dprod a.
\end{equation}
The definitions of inner and outer products can be extended to
general multivectors
\begin{eqnarray}
    a \dprod b &=& \sum_{\alpha,\beta} \left<<\!a\!>_\alpha \;
    <\!b\!>_\beta \right>_{|\alpha-\beta|},\\
    a \wprod b &=& \sum_{\alpha,\beta} \left<<\!a\!>_\alpha \;
    <\!b\!>_\beta \right>_{\alpha+\beta}.
\end{eqnarray}
Two other useful products are the scalar product, denoted as
$<\!ab\!>$ and commutator product, defined by
\begin{equation}
    a \times b  = \sfrac{1}{2}(ab - ba).
\end{equation}
In mixed product expressions we will use the convention that inner
and outer products take precedence over geometric products as said
above.

We will encounter exponentials with multivector exponents; two
particular cases of exponentiation are specially important. If $u$
is such that $u^2 = -1$ and $\theta$ is a scalar
\begin{equation}
\begin{split}
   \mathrm{e}^{u \theta} &= 1 + u \theta -\frac{\theta^2}{2!} - u
    \frac{\theta^3}{3!} + \frac{\theta^4}{4!} + \ldots   \\
    &= 1 - \frac{\theta^2}{2!} +\frac{\theta^4}{4!}- \ldots \{=
    \cos \theta \}  \\
    &\quad + u \theta - u \frac{\theta^3}{3!} + \ldots \{= u \sin
    \theta\} \\
    &=  \cos \theta + u \sin \theta.
\end{split}
\end{equation}
Conversely if $h$ is such that $h^2 =1$
\begin{equation}
\begin{split}
    \mathrm{e}^{h \theta} &= 1 + h \theta +\frac{\theta^2}{2!} + h
    \frac{\theta^3}{3!} + \frac{\theta^4}{4!} + \ldots   \\
    &= 1 + \frac{\theta^2}{2!} +\frac{\theta^4}{4!}+ \ldots \{=
    \cosh \theta \}  \\
    &\quad + h \theta + h \frac{\theta^3}{3!} + \ldots \{= h \sinh
    \theta\}  \\
    &=  \cosh \theta + h \sinh \theta. \nonumber
\end{split}
\end{equation}
The exponential of bivectors is useful for defining rotations; a
rotation of vector $a$ by angle $\theta$ on the $\sigma_{12}$ plane
is performed by
\begin{equation}
    a' = \mathrm{e}^{\sigma_{21} \theta/2} a
    \mathrm{e}^{\sigma_{12} \theta/2}= \tilde{R} a R;
\end{equation}
the tilde denotes reversion and reverses the order of all products.
As a check we make $a = \sigma_1$
\begin{equation}
\begin{split}
    \mathrm{e}^{-\sigma_{12} \theta/2} \sigma_1
    \mathrm{e}^{\sigma_{12} \theta/2} &=
    \left(\cos \frac{\theta}{2} - \sigma_{12}
    \sin \frac{\theta}{2}\right) \sigma_1  \\
    &\quad \ast \left(\cos \frac{\theta}{2} + \sigma_{12} \sin
    \frac{\theta}{2}\right)  \\
    &= \cos \theta \sigma_1 + \sin \theta \sigma_2.
\end{split}
\end{equation}
Similarly, if we had made $a = \sigma_2,$ the result would have been
$-\sin \theta \sigma_1 + \cos \theta \sigma_2.$

If we use $B$ to represent a bivector whose plane is normal to
$\sigma_0$ and define its norm by $|B| = (B \tilde{B})^{1/2},$ a
general rotation in 4-space is represented by the rotor
\begin{equation}
    R \equiv e^{-B/2} = \cos\left(\frac{|B|}{2}\right) -  \frac{B}{|B|}
    \sin\left(\frac{|B|}{2}\right).
\end{equation}
The rotation angle is $|B|$ and the rotation plane is defined by
$B.$ A rotor is defined as a unitary multivector verifying
$\tilde{R}R =1$; we are particularly interested in rotors with
bivector components. It is more general to define a rotation by a
plane (bivector) then by an axis (vector) because the latter only
works in 3D while the former is applicable in any dimension. When
the plane of bivector $B$ contains $\sigma_0$, a similar operation
does not produce a rotation but produces a boost instead. Take for
instance $B = \sigma_{01} \theta/2$ and define the transformation
operator $T = \exp( B)$; a transformation of the basis vector
$\sigma_0$ produces
\begin{equation}
\begin{split}
    a' &= \tilde{T} \sigma_0 T = \mathrm{e}^{-\sigma_{01}\theta/2}
    \sigma_0 \mathrm{e}^{\sigma_{01}\theta/2}  \\
    &= \left(\cosh \frac{\theta}{2} - \sigma_{01}
    \sinh \frac{\theta}{2}\right) \sigma_0  \\
    &\quad \ast\left(\cosh \frac{\theta}{2} + \sigma_{01} \sinh
    \frac{\theta}{2}\right) \\
    &= \cosh \theta \sigma_0 + \sinh \theta \sigma_1.
\end{split}
\end{equation}

\section{Reciprocal frame and derivative operators \label{derivatives}}

A reciprocal frame is defined by the condition
\begin{equation}
    \label{eq:recframe}
    g^\alpha \dprod g_\beta = {\delta^\alpha}_\beta.
\end{equation}
Defining $g^{\alpha \beta}$ as the inverse of $g_{\alpha \beta}$,
the matrix product of the two must be the identity matrix, which we
can state as
\begin{equation}
    g^{\alpha \gamma} g_{\beta \gamma} = {\delta^\alpha}_\beta.
\end{equation}
Using the definition (\ref{eq:metrictens}) we have
\begin{equation}
    \left(g^{\alpha \gamma} g_\gamma \right)\dprod g_\beta =
    {\delta^\alpha}_\beta;
\end{equation}
comparing with Eq.\ (\ref{eq:recframe}) we  determine $g^\alpha$
\begin{equation}
    g^\alpha = g^{\alpha \gamma} g_\gamma.
\end{equation}
It would be easy to verify that it is also $g^{\alpha \beta} =
g^\alpha \cdot g^\beta$ and $g_\alpha = g_{\alpha \gamma} g^\gamma$.

In many situations of great interest the frame vectors $g_\alpha$
can be expressed in terms of an orthonormed frame given by Eqs.\
(\ref{eq:basis}). If the frame vectors can be expressed as linear
combination of the orthonormed ones we have
\begin{equation}
    \label{eq:indexframe}
    g_\alpha = {n^\beta}_\alpha \sigma_\beta,
\end{equation}
where ${n^\beta}_\alpha$ is called the \emph{refractive index
tensor} or simply the \emph{refractive index} as said in the main
text. When the refractive index is the identity we have $g_\alpha =
\sigma_\alpha$ for the main or direct frame and $g^0 = -\sigma_0$,
$g^i = \sigma_i$ for the reciprocal frame, so that Eq.\
(\ref{eq:recframe}) is verified.

The first use we will make of the reciprocal frame is for the
definition of two derivative operators. In flat space we define the
vector derivative
\begin{equation}
    \nabla = \sigma^\alpha\partial_\alpha.
\end{equation}
It will be convenient, sometimes, to use vector derivatives in
subspaces of 5D space; these will be denoted by an upper index
before the $\nabla$ and the particular index used determines the
subspace to which the derivative applies; For instance $^m\nabla =
\sigma^m \partial_m = \sigma^1 \partial_1 + \sigma^2 \partial_2 +
\sigma^3 \partial_3.$ In 5-dimensional space it will be useful to
split the vector derivative into its time and 4-dimensional parts
\begin{equation}
    \nabla = -\sigma_0\partial_t + \sigma^i \partial_i
    = -\sigma_0\partial_t
    + \pre{i}\nabla.
\end{equation}
Consistently with the boldface notation for 3-dimensional vectors
$\pre{m}\nabla$ will be denoted by $\bm{\nabla}$. We will use over
arrows, when necessary, to imply that the vector derivative is
applied to a function which is not immediately on its right; for
instance in $\overrightarrow{\nabla}A \overleftarrow{B}$ and in
$\overrightarrow{B}A \overleftarrow{\nabla}$ the derivative operator
is applied to function $B$.

The second derivative operator is called covariant derivative,
sometimes also designated by \emph{Dirac operator}, and it is
defined with recourse to the reciprocal frame $g^\alpha$
\begin{equation}
    \mathrm{D} = g^\alpha\partial_\alpha.
\end{equation}
Taking into account the definition of the reciprocal frame
(\ref{eq:recframe}) we see that the covariant derivative is also a
vector. In cases where there is a refractive index, it will  be
possible to define both derivatives in the same space.

Vector derivatives can also be left or right multiplied with other
vectors or multivectors. For instance, when $\nabla$ is multiplied
by vector $a$ on the right the result comprises scalar and bivector
terms $\nabla a = \nabla \cdot a + \nabla \wprod a.$ The scalar part
can be immediately associated with the divergence and the bivector
part is called the exterior derivative; in the particular case of
Euclidean 3-dimensional space it is possible to define the
$\mathrm{curl}$ of a vector by $\mathrm{curl}(\mathbf{a}) =
-\sigma_{123}\bm{\nabla} \wprod \mathbf{a}.$

We define also second order differential operators, designated
Laplacian and covariant Laplacian respectively, resulting from the
inner product of one derivative operator by itself. The square of a
vector is always a scalar and the vector derivative is no exception,
so the Laplacian is a scalar operator, which consequently acts
separately in each component of a multivector. For $4+1$ flat space
it is
\begin{equation}
    \nabla^2 = -\frac{\partial^2}{\partial t^2} + \pre{i}\nabla^2.
\end{equation}
In Minkowski spacetime it is usual to call D'Alembertian to
$\nabla^2$ but we prefer to use the Laplacian designation for all
spaces; this is consistent with the definition of tensor Laplacian
found, for instance, in \cite{Arfken95}. One sees immediately that a
4-dimensional wave equation is obtained zeroing the Laplacian of
some function
\begin{equation}
    \label{eq:4dwave}
    \nabla^2 \Psi = \left(-\frac{\partial^2}{\partial t^2} +
    \pre{i}\nabla^2\right)\Psi = 0.
\end{equation}
This procedure was used in Ref.\ \cite{Almeida05:4} for the
derivation of special relativity and  extended in Ref.\
\cite{Almeida06:2} to general curved spaces.

\end{appendix}


\end{document}